\documentclass[journal]{IEEEtai}

\usepackage[colorlinks,urlcolor=blue,linkcolor=blue,citecolor=blue]{hyperref}
\usepackage{xcolor}
\usepackage{color,array}
\usepackage{comment}
\usepackage{multirow}
\usepackage{accents}
\usepackage{algorithm}
\usepackage{algpseudocode}
\usepackage{graphicx}

\usepackage{cite}

\begin{document}

%\title{Emotion-Driven Audio-Visual Speech Enhancement for Assistive Hearing} 
%\title{Enhancing Audio-Visual Communication in Noisy Environments via Emotion-Based Contextual Cues} 

%\title{Enhancing Audio-Visual Communication in Noisy Environments via Emotion-Based Contextual Cues} 
\title{Audio-Visual Speech Enhancement in Noisy Environments via Emotion-Based Contextual Cues}

\author{Tassadaq Hussain, Kia Dashtipour, Yu Tsao, and Amir Hussain 

%\thanks{This paragraph of the first footnote will contain the date on which you submitted your paper for review. It will also contain support information, including sponsor and financial support acknowledgement. For example, ``This work was supported in part by the U.S. Department of Commerce under Grant BS123456.'' }

%\thanks{The next few paragraphs should contain the authors' current affiliations, including current address and e-mail. For example, F. A. Author is with the National Institute of Standards and Technology, Boulder, CO 80305 USA (e-mail: author@boulder.nist.gov).}

\thanks{Tassadaq Hussain is with the School of Computing, Edinburgh Napier University, UK.  (e-mail: t.hussain@napier.ac.uk).}
\thanks{Kia Dashtipour is with the School of Computing, Edinburgh Napier University, UK. (e-mail: k.dashtipour@napier.ac.uk).}
\thanks{Yu Tsao is with the Center for Information Technology and Innovation, Academia Sinica, Taiwan.  (e-mail: yu.tsao@citi.sinica.edu.tw).}
\thanks{Amir Hussain is with the School of Computing, Edinburgh Napier University, UK. (e-mail: a.hussain@napier.ac.uk).}
\thanks{This paragraph will include the Associate Editor who handled your paper.}}

%\markboth{Journal of IEEE Transactions on Artificial Intelligence, Vol. 00, No. 0, Month 2020}
%{First A. Author \MakeLowercase{\textit{et al.}}: Bare Demo of IEEEtai.cls for IEEE Journals of IEEE Transactions on Artificial Intelligence}

\maketitle

\begin{abstract}
\textcolor{blue}{
In real-world environments, background noise significantly degrades the intelligibility and clarity of human speech. Audio-visual speech enhancement (AVSE) attempts to restore speech quality, but existing methods often fall short, particularly in dynamic noise conditions. This study investigates the inclusion of emotion as a novel contextual cue within AVSE, hypothesizing that incorporating emotional understanding can improve speech enhancement performance. We propose a novel emotion-aware AVSE system that leverages both auditory and visual information. It extracts emotional features from the facial landmarks of the speaker and fuses them with corresponding audio and visual modalities. This enriched data serves as input to a deep UNet-based encoder-decoder network, specifically designed to orchestrate the fusion of multimodal information enhanced with emotion. The network iteratively refines the enhanced speech representation through an encoder-decoder architecture, guided by perceptually-inspired loss functions for joint learning and optimization. We train and evaluate the model on the CMU Multimodal Opinion Sentiment and Emotion Intensity (CMU-MOSEI) dataset, a rich repository of audio-visual recordings with annotated emotions, specifically chosen for its diversity and real-world scenarios. Our comprehensive evaluation demonstrates the effectiveness of emotion as a contextual cue for AVSE. By integrating emotional features, the proposed system achieves significant improvements in both objective and subjective assessments of speech quality and intelligibility, especially in challenging noise environments. Compared to baseline AVSE and audio-only speech enhancement systems, our approach exhibits a noticeable increase in PESQ (from 1.43 to 1.67) and STOI (from 0.70 to 0.74), indicating higher perceptual quality and intelligibility. Large-scale listening tests corroborate these findings, suggesting improved human understanding of enhanced speech.
}
\begin{comment}{
Our research builds upon a deep U-Net-based encoder decoder network, orchestrating the fusion of auditory and visual information to enhance speech quality and intelligibility. By incorporating visual cues, our model transcends traditional AVSE approaches by honing in on the auditory characteristics of specific speakers within a given scene. To train our comprehensive AVSE model, we harness the CMU Multimodal Opinion Sentiment and Emotion Intensity (CMU-MOSEI) dataset, an extensive repository of audio-video recordings with emotion labels. This dataset encapsulates hundreds of hours of audio-visual data, enabling us to fortify our model's ability to handle real-world challenges.
Our findings showcase the remarkable efficacy of our AVSE system, which excels not only in tackling conventional speech enhancement hurdles but also in surmounting real-world scenarios fraught with speech and noise interference. This includes dynamic, non-stationary noise environments where our model consistently outperforms previous AVSE systems. Our innovative approach is poised to transform audio-visual speech enhancement by introducing a novel dimension: emotion awareness. By integrating the embedding of emotion frameworks as contextual cues, we anticipate further advancements in speech quality and intelligibility, enriching communication experiences in noisy environments.}
\end{comment}

\end{abstract}

\begin{comment}{
\begin{IEEEImpStatement}
Our research at the intersection of audio-visual speech enhancement and emotion integration holds immense promise. Beyond enhancing speech quality for individuals with hearing impairment and hearing aid users, this innovation has transformative implications for diverse fields. It revolutionizes assistive technologies, empowering those with hearing challenges to engage seamlessly in conversations. Moreover, it fosters advancements in conversational AI, making automated systems more attuned to human emotions and intent. Telecommunication systems benefit from heightened clarity and understanding in noisy environments, enhancing the user experience. This research transcends conventional boundaries and propels human-computer interaction to new heights, enriching the quality of communication in real-world scenarios. As we embrace emotion-based contextual cues, we embark on a journey toward a more inclusive, empathetic, and accessible world for all.
\end{IEEEImpStatement}
\end{comment}


\section{Introduction}
\IEEEPARstart{S}{peech} enhancement (SE) is a critical area in signal processing, primarily driven by the imperative to improve the quality and intelligibility of spoken communication. Despite extensive efforts in this field, conventional SE algorithms have often fallen short in effectively addressing the challenges posed by real-world noisy environments. Traditional techniques frequently struggle to disentangle desired speech from background noise, especially in the presence of non-stationary or highly unpredictable acoustic interferences. Their limited capacity to differentiate between speech and noise sources has led to a quest for more robust solutions \cite{ephraim1984speech} \cite{martin2001noise} \cite{paliwal2012speech} \cite{chern2023audio}.

\textcolor{blue}{
The emergence of deep learning (DL)-based approaches has revolutionized the landscape of SE \cite{lu2013speech} \cite{xu2014regression}. These methodologies leverage the power of neural networks to learn complex patterns and relationships, enabling them to surpass their conventional counterparts. DL models, particularly convolutional and recurrent neural networks, have displayed a remarkable capability to enhance the quality and intelligibility of speech in noisy conditions \cite{zhang2022multi} \cite{zhao2022frcrn} \cite{zhang2022multi}. However, despite the significant advancements in DL-based SE, the challenges posed by specific scenarios, such as the presence of multiple speakers and challenging acoustic environments, remained unaddressed. This limitation prompted the exploration of audio-visual speech enhancement (AVSE) systems, which harness both auditory and visual cues to enhance speech signals. 
}
\textcolor{blue}{
The integration of visual information, such as lip movements and facial landmarks, alongside audio data, has offered promising results in mitigating the impact of noise and enhancing the quality of spoken communication \cite{afouras2018deep}. AVSE systems, by jointly processing auditory and visual signals, can provide superior performance in scenarios where traditional audio-only approaches fall short, marking a significant step forward in the pursuit of intelligible and noise-resistant speech \cite{afouras2018conversation} \cite{sadeghi2020audio} \cite{michelsanti2021overview} \cite{chuang2022improved} \cite{golmakani2023audio}. Similarly, researchers have taken substantial steps to enhance the performance of both Audio-only SE (ASE) and AVSE systems, especially in the presence of extreme noise conditions. An innovative approach involves the utilization of distinct loss functions, particularly distance-based and perceptually-inspired ones, to optimize speech quality and intelligibility. These loss functions take into account perceptual aspects of speech quality  \cite{kolbaek2020loss}, notably the Perceptual Evaluation of Speech Quality (PESQ) and Short-Time Objective Intelligibility (STOI), and are seamlessly integrated into the training process \cite{hsieh2020improving} \cite{li2020imetricgan} \cite{hussain2021towards} \cite{hussain2022speech}. This integration ensures that the ASE and AVSE system closely aligns with human auditory perception, resulting in a marked improvement in speech signal quality and intelligibility, even in the face of challenging and dynamic noise environments.
}
\begin{comment} {
    
More recently, researchers have combined the distinct properties of a regression and a classification system to achieve remarkable performances. For example, the current trend involves utilizing speech enhancement systems as pre-processing blocks to suppress noise effectively and enhance the accuracy of emotion recognition systems \cite{triantafyllopoulos2019towards} \cite{goncalves2022robust}. By using regression systems (SE) to pre-process the data for classification systems (emotion recognition), researchers have created more robust and context-aware solutions for affective computing \cite{chen2023noise} \cite{tiwari2020multi}.
}
\end{comment}
\textcolor{blue}{
In recent years, a notable paradigm shift has been observed in the domain of affective computing. Researchers have explored innovative approaches that effectively leverage both regression and classification systems to achieve remarkable performance enhancements. A prominent trend in this context involves the integration of SE systems as crucial pre-processing components to significantly mitigate noise interference, consequently leading to a substantial improvement in the accuracy of emotion recognition systems. One significant development in this direction is the combined use of regression systems, particularly SE models, as preliminary stages in the pipeline for classification systems focused on emotion recognition. This novel approach has demonstrated exceptional potential and garnered substantial attention in the research community. By employing SE as a pre-processing step, researchers aim to create highly robust and context-aware solutions for affective computing, particularly in challenging, real-world scenarios where environmental noise can significantly affect the accuracy of emotion detection. For example, this strategy is exemplified in recent studies such as Triantafyllopoulos et al. \cite{triantafyllopoulos2019towards}, who introduced a novel framework that effectively utilizes SE techniques to preprocess audio data before its classification for emotion recognition. Their research showcases the potential of SE as a valuable tool for enhancing emotion detection accuracy in noisy environments. Additionally, Michael et al. \cite{neumann2021investigations} explored a robust approach for AV emotion recognition under noisy acoustic conditions with a focus on speech features. Furthermore, the study by Chen et al. delves into the intricacies of combining SE and emotion recognition. Their work emphasizes the essential role of SE as a pre-processing mechanism to ensure the robustness of emotion recognition systems, particularly in contexts where audio quality may be compromised by environmental noise \cite{chen2023noise}. In a similar vein, Tiwari et al. contributed to the research landscape by investigating multi-modal approaches that integrate SE and emotion recognition, emphasizing the potential for context-aware affective computing \cite{tiwari2020multi}.
}
\textcolor{blue}{
Motivated by the growing synergy between SE and emotion recognition systems, as well as the need for more robust and context-aware SE solutions. We propose a novel SE approach that harnesses emotion recognition as a contextual cue to improve performance in extremely noisy environments. Our approach integrates the emotion recognition mechanism into the SE process to enhance the quality of audio signals and effectively suppress noise. By integrating emotion recognition into the AVSE process, we unlock the potential to adapt and fine-tune the enhancement procedure according to the speaker's emotional state. This adaptation can lead to context-aware and emotionally relevant SE, ensuring that not only background noise is reduced, but the speech itself remains natural and emotionally resonant. In noisy conditions, where traditional enhancement methods may falter, this fusion of emotion and SE becomes especially crucial in achieving enhanced speech quality and intelligibility, providing more authentic and context-aware communication experiences. This novel approach has the potential to transform the landscape of SE and affective computing by addressing real-world challenges where emotional nuances coexist with acoustic disturbances. Our work is significant because it offers a promising paradigm shift in AVSE. The experimental results demonstrate invaluable insights and open the doors to the development of future emotion-aware applications across diverse domains. Our work not only contributes to advancing the current state of research but also holds the promise of positively impacting various facets of human-machine interaction and communication i.e., conversational AI. The main contributions of our paper are; i) we introduce a joint model for AV fusion with emotion features integration to exploit the complementary relationship across modalities for emotional nuances coexisting with acoustic disturbances. ii) Unlike previous approaches, our model simultaneously leverages intra- (temporal, spectral, and spatial relationship) and intermodal (correlation and synchronisation between the audio and video signals) relationships to capture complementary relationships effectively. iii) By integrating emotional cues and deploying the joint feature representation, our model also reduces heterogeneity across audio and visual features, resulting in robust AV feature representations.iv) Extensive experiments on the challenging MOSEI dataset demonstrate that our proposed fusion model outperforms state-of-the-art AVSE models. Spectrogram analysis shows that our model can efficiently leverage complementary contextual relationships while retaining intramodal relationships.
}

\begin{figure*}[!t]
    \centering
    \includegraphics[width=\linewidth]{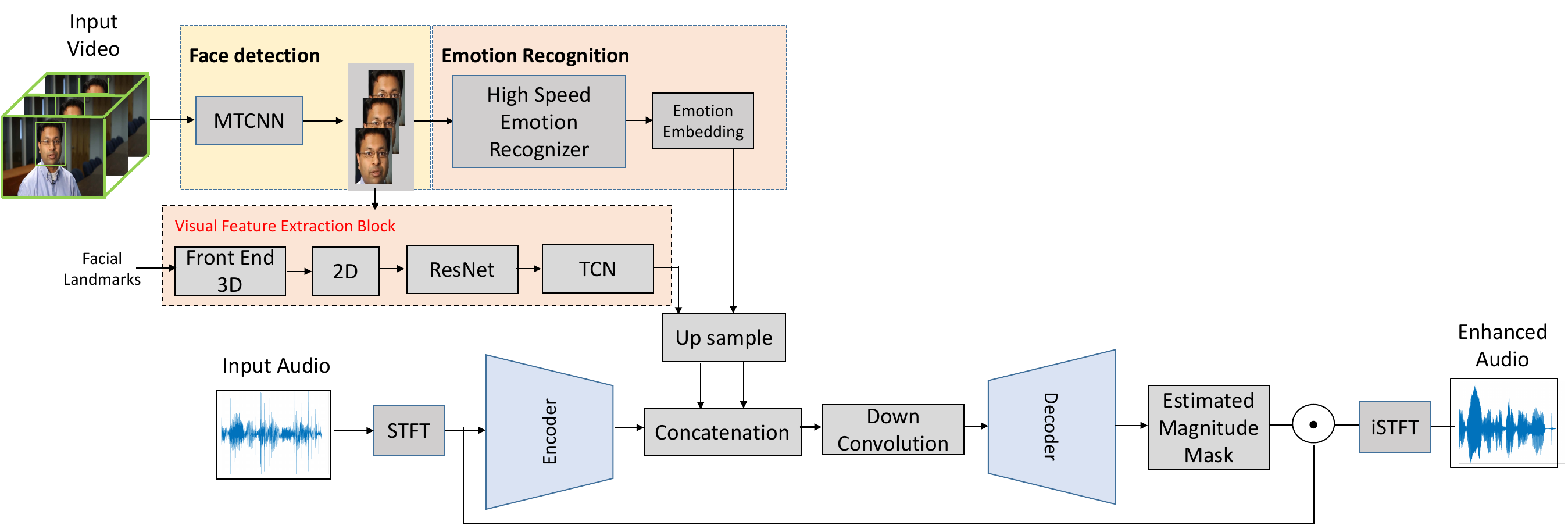}
    \caption{Architecture of the proposed Emotion-AVSE system.} 
\end{figure*}
%########################

%\section{Proposed Emotion-aware AVSE System}
\section{Utilizing Emotion Cues for AVSE System}
\textcolor{blue}{
We develop an encoder-decoder-style UNet architecture to train AVSE models. Figure 1 provides an overview of our proposed emotion-aware AVSE system architecture. Algorithm 1 delineates the stepwise process of our novel AVSE algorithm, designed to enhance speech quality in noisy environments through the integration of facial information and emotion-based contextual cues. The proposed algorithm represents a comprehensive approach to AVSE, integrating facial information and emotional cues. It begins with the processing of a noisy speech signal through Short-Time Fourier Transform (STFT) for audio feature extraction. Subsequently, facial detection using Multi-task Convolutional Neural Network (MTCNN) \cite{zhang2016joint} is applied to identify facial regions in video frames. Visual feature extraction involves spatiotemporal convolution and ResNet architectures, while emotion features are extracted using an EfficientNet model\cite{tan2019efficientnet}. The three modalities, audio, visual, and emotion, are then seamlessly fused using a concatenation operation. The resulting composite tensor undergoes refinement through downsampling convolutional layers before being processed by a UNet decoder to generate an enhanced spectrogram. The enhanced speech signal is synthesized by combining the enhanced spectrogram with the phase of the original noisy speech. This algorithm showcases a robust framework for enhancing speech quality by integrating emotion-aware visual information into the audio processing pipeline.
}

\subsubsection{Audio feature extraction}
\textcolor{blue}{
For audio processing, we send the magnitude of $F \times T$ dimensional noisy STFT to the network (where $F$ and $T$ are frequency and time dimensions) such that 
\begin{equation}
X_{i} = STFT(x_{noisy})
\end{equation}
Here, we denote $X_{i}$ as the magnitude spectrum of the $i$-th noisy speech signal $x_{noisy}$. The initial audio data is subsequently processed through two convolutional layers with a filter size of $4$ and a stride of $2$, resulting in reduced time-frequency dimensions 
\begin{equation}
X_{conv} = [Conv2D(X_{i})]*2
\end{equation}
where $X_{conv}$ represents the data after the first and second convolution and Conv2D represents the 2D convolution operation with the filter size of $4$ and stride of $2$. Following this initial processing, the data then undergoes further refinement through three consecutive convolutional blocks and subsequent frequency pooling layers. Each convolutional layers employs a filter of size $3$ and a stride of $1$. Subsequently, the frequency dimension is halved due to the subsequent frequency pooling layers. These convolutional and subsequent frequency poling layers are responsible for down-sampling the features, ultimately arriving at their outputs, represented as $X_{blocks}$.This entire process, encapsulated by the convolutional blocks and frequency pooling operations, is detailed below.
\begin{equation}
X_{blocks} = [FrequencyPooling(ConvBlocks(X_{conv}))] * 3
\end{equation}
Note that throughout the convolutional block processing, the spatial dimension is preserved. The output of the final convolutional block is designated as $A_{i}$, representing the data at the frequency pooling layer for the $i$-th frame in the above equation.
\begin{equation}
A_{i} = X_{blocks}
\end{equation}
}

\begin{algorithm}
%\caption{Audio-Visual Speech Enhancement Algorithm}\label{alg:avse}
\caption{Stepwise AVSE Algorithm in Noisy Environments with Facial Information and Emotion-Based Contextual Cues}\label{alg:avse}
\begin{algorithmic}[1]
\Require Noisy speech signal $x_{noisy}$, Visual data $v_{video}$
\Ensure Enhanced speech signal $x_{enhanced}$ \\

\textbf{Audio Processing:}
\State \hspace{1em} $X_{i} \gets {Magnitude spectrum} (x_{noisy})$
%\State $X_{\text{conv1}}, X_{\text{conv2}} \gets \text{Conv2D}(X_{\text{audio}}, \text{filter\_size}=4, \text{stride}=2)$
%\State $X_{\text{block1}}, X_{\text{block2}}, X_{\text{block3}} \gets \text{FrequencyPooling}(\text{ConvBlocks}(X_{\text{conv2}}))$
\State \hspace{1em} $A_{i} \gets {UNetAudioEncoder(X_{i}))}$

\State \textbf{Visual Processing:}
\State \hspace{1em} $I_{i} \gets {VideoReader}(video)$
\State \hspace{1em} $D_{i} \gets {MTCNN}(I_{i})$
\State \hspace{1em} $C_{conv_{i}} \gets ResNet(D_{i})$
\State \hspace{1em} $T_{i} \gets TCN(C_{conv_{i}})$
\State \hspace{1em} $V_{i} \gets Upsample(T_{i})$

\State \textbf{Emotion Processing:}
% Connection Line
\State \textbf{Connection:} \Comment{Connect Visual Processing to Emotion Processing}
\State \hspace{0.5cm} $E_{EffNet} \gets {EfficientNet}(D_{i})$ \Comment{Use the output of MTCNN in Emotion Processing}
%\State $E_{i} \gets {EfficientNet}(D_{i})$
\State \hspace{1em} $E_{linear} \gets {LinearLayer}(E_{EffNet})$
\State \hspace{1em} $E_{i} \gets {Upsample}(E_{linear})$

\State \textbf{Integration and Decoding:}
\State \hspace{1em} $M_{composite} \gets {Concatenate}(A_{i}, V_{i}, E_{i})$
\State \hspace{1em} $M_{composite_{down}} \gets {Downsample}(M_{composite})$
\State \hspace{1em} $S_{enhanced} \gets {UNetDecoder}(M_{composite_{down}})$
\State \hspace{1em} $x_{enhanced} \gets {iSTFT}(S_{enhanced}, Phase(x_{noisy}))$
\end{algorithmic}
\end{algorithm}
\subsubsection{Face Detection and Facial Emotion Emotion Recognition}
\textcolor{blue}{
In our approach, we employ pre-trained models to facilitate face detection and Facial Emotion Recognition (FER). The initial step in this pipeline is face detection, wherein we utilize the MTCNN model for robust and efficient face localization from video streams. MTCNN excels at identifying faces within images or video frames, functioning independently for each frame in the latter scenario. MTCNN identifies and isolates facial regions within the video frames.
}
\begin{comment}
In the initial stage, MTCNN leverages a fully convolutional network to identify candidate windows and calculate bounding box regression vectors. This process closely follows established methodologies. The bounding box regression vectors are subsequently utilized to enhance the quality of the candidates. Non-maximum suppression (NMS) is then applied to consolidate highly overlapping candidates. The second stage involves all candidates undergoing evaluation within another Convolutional Neural Network to refine the pool of candidates by eliminating false positives, performing calibration through bounding box regression, and further employing NMS to merge candidate windows effectively. The third and final stage mirrors the structure of the second stage but expands its scope to provide a more comprehensive description of facial features. In this stage, the network is equipped to output the positions of five facial landmarks (left eye, right eye, nose, left mouth corner, and right mouth corner), enhancing its capabilities. These facial regions are then provided as input to the following stages of our pipeline, where facial emotions are subsequently recognized.
\end{comment}
\textcolor{blue}{
The task of face detection in MOSEI YouTube videos is confronted with multiple complexities due to differences in video durations and frame sizes. To address these intricacies, we adopt a standardized approach involving 4-second video segments. Utilizing the VideoReader function, we extract individual frames ($I_{i}$) from the video such that
\begin{equation}
I_{i} = VideoReader(video)
\end{equation}  
Subsequently, the MTCNN model is employed to perform face detection on these visual frames. MTCNN identifies faces and provides the precise positions of five key facial landmarks: the left eye, right eye, nose, left mouth corner, and right mouth corner. The face localization process for the $i$-th frame is expressed as
\begin{equation}
D_{i} = MTCNN(I_{i})
\end{equation}  
with $D_{i}$ symbolizing the resulting detected bounding box. The outcome of MTCNN is a structured frame format characterized by dimensions (F, C, H, W), where F pertains to the number of visual frames, C designates the channel, and H and W correspond to height and width. To ensure consistency, all facial images are uniformly resized to dimensions of 224x224.
}
\textcolor{blue}{
The process of detecting faces is followed by their subsequent analysis through the visual feature extraction and Facial Emotion Recognition (FER) blocks. The visual feature extraction pipeline is initiated by a 3D convolutional layer designed for spatiotemporal convolution, complemented by an 18-layer ResNet-18 architecture, as described by He et al. (2016) \cite{he2016deep}. Within the 3D convolutional layer, short-term dynamics related to lip articulations are effectively captured. This is accomplished through the application of a convolutional layer featuring 64 3D kernels with dimensions of $5\times7\times7$ (time$\times$width$\times$height) and a stride of $1\times2\times2$. The entire operation can be succinctly denoted as
\begin{equation}
C_{conv_{i}} = ResNet(Conv3D(D_{i}), 64, (5, 7, 7))
\end{equation}  
}
\textcolor{blue}{
Subsequently, the output of the ResNet-18 architecture feeds into a temporal convolutional network (TCN), following the methodology outlined in Martinez et al.'s work on lip-reading \cite{martinez2020lipreading}. This network is provided with a temporal sequence of face images, each of size $N \times 224 \times 224$, where $N$ represents the number of frames. The output of the visual feature network yields a 512-D vector for each image frame. This process can be formally expressed as 
\begin{equation}
T_{i} = TCN(C_{conv_{i}})
\end{equation} 
To ensure compatibility with the audio features, the visual features are upsampled to match the audio features' sampling rate, denoted as
\begin{equation}
V_{i} = Upsample(T_{i})
\end{equation} 
}
\textcolor{blue}{
In parallel, the Facial Emotion Recognition (FER) stage is executed through the deployment of the High Speech Emotion Recognition (HSEmotion) model \cite{savchenko2022hsemotion}. HSEmotion is a versatile tool capable of processing static images or facial videos, offering a comprehensive analysis of emotional expressions. It produces two distinct outputs: high-dimensional visual embeddings, recognized as "emotional features," and posterior probabilities associated with eight distinct emotions: Anger, Contempt, Disgust, Fear, Happiness, Neutral, Sadness, and Surprise. These outputs are captured through the EfficientNet model \cite{tan2019efficientnet} such as,
\begin{equation}
E_{EffNet} = EfficientNet(D_{i})
\end{equation}
where $E_{EffNet}$ signifies the high-dimensional visual embedding or emotional features derived from the EfficientNet model.
}
\textcolor{blue}{
It is worth noting that, in this preliminary exploration, no fine-tuning was performed on the architecture or parameters of the EfficientNet model explicitly for the task of emotion recognition from facial expressions. Instead, the model was harnessed to extract emotional feature representations from the faces detected by the MTCNN model. This unique approach allowed us to investigate the potential of employing these emotional feature representations as contextual cues to enhance the overall system's performance, with a particular emphasis on emotion recognition. The EfficientNet model excels at capturing subtle emotional nuances within facial images, thereby furnishing a comprehensive and valuable representation of the emotional states conveyed within these images.
}

\subsubsection{Multimodal fusion and speech resynthesis}
\textcolor{blue}{
The output of the FER model yields emotional features presented in the form of a tensor with dimensions $(B, 1280, F)$, where $B$ designates the batch size, 1280 represents the emotional embedding dimension, and $F$ accounts for the number of visual frames in a video sequence. To ensure harmonization across the modalities, the emotional features undergo transformation via a linear layer, culminating in feature embeddings characterized by dimensions $(B, 512, F)$. This transformation is represented as
\begin{equation}
E_{linear} = LinearLayer(E_{EffNet})
\end{equation}
Following the linear transformation, the output of the linear layer is upsampled, converting it into a tensor with dimensions $(B, 512, 8, F)$. This ensures alignment with the output from the audio UNet encoder and the visual feature network. The upsampled emotional features are denoted as 
\begin{equation}
E_{i} = Upsample(E_{linear})
\end{equation}
Subsequently, the final integration stage takes shape, uniting the outputs from the audio encoder, visual feature network, and emotion network, resulting in an embedding tensor characterized by dimensions $(B, 1536, 8, F)$. This integration process is concisely represented as
\begin{equation}
M_{composite} = Concatenate(A_{i}, V_{i}, E_{i})
\end{equation}
The composite tensor, $M_{composite}$, then proceeds through additional processing via a downsampling convolutional layer, yielding an output tensor with dimensions $(B, 512, 8, F)$, and formalized as 
\begin{equation}
M_{{composite}_{down}} = Downsample(M_{composite})
\end{equation}
Ultimately, the processed output is fed into the UNet decoder, a pivotal step in the generation of the final enhanced spectrogram and can be written as
\begin{equation}
S_{enhanced} = UNetDecoder(M_{{composite}_{down}})
\end{equation}
}
\textcolor{blue}{
In this work, we employed the "enet\_b0\_8\_va\_mtl" model of EffientNet, noteworthy for extending its predictions beyond the eight fundamental facial expressions to encompass valence and arousal. These additional emotional dimensions, concerned with assessing the level of emotional intensity and the positive or negative aspects of human behavior, exceed the current scope of this paper. Nevertheless, we anticipate integrating valence and arousal dimensions into our future research, thereby enhancing the comprehensiveness of our emotion-aware SE system.
}
\section{Experimental Setup}
\subsection{Description of the CMU-MOSEI Dataset}
\textcolor{blue}{
The CMU Multimodal Opinion Sentiment and Emotion Intensity (MOSEI) dataset stands out for its richness and suitability for various research tasks. Comprising over 1,000 YouTube videos, MOSEI captures diverse and realistic emotional expressions from multiple speakers across a wide range of sentiments. Notably, the dataset provides detailed emotion and sentiment labels, making it ideal for training emotion recognition, sentiment analysis, and opinion mining models. The use of opinionated audio-visual content ensures the data reflects real-world scenarios and the complex nuances of human emotions. MOSEI's significant diversity in both speakers and emotional expression guarantees its generalizability and effectiveness in various research applications.
}
\begin{comment}{
The CMU Multimodal Opinion Sentiment and Emotion Intensity (MOSEI) dataset comprises a substantial number of utterances, segments, and other data points, making it a rich and extensive resource for research. MOSEI contains data from over 1,000 YouTube videos, which are the source of its diverse and realistic content 
%These videos yield approximately 23,500 segments, with each segment typically corresponding to a short duration of speech, 
thus making it suitable for various research tasks, including sentiment and emotion analysis. The dataset offers a comprehensive view of emotional expressions, encompassing multiple speakers, and it covers a wide range of emotions and sentiments. This diverse dataset boasts detailed emotion and sentiment labels, making it an ideal choice for training models aimed at recognizing emotions, sentiments, and opinions. The data is derived from opinionated AV content, a context frequently encountered in real-world communication, and thus, offers a faithful representation of the types of emotional expressions one may encounter in daily life. MOSEI's notable diversity encompasses a wide spectrum of emotions and sentiments, ensuring that the data is reflective of the complexity and richness of human emotions. %Moreover, the dataset is structured to provide information across various modalities, including audio, visual, and text, enhancing its utility for researchers in the field of emotion-aware audio-visual speech enhancement.
}
\end{comment}

\subsection{Analyzing Audio-Visual Recordings with Emotions}
\textcolor{blue}{
The MOSEI dataset offers predefined train, validation, and test splits for robust model development. The training set encompasses a diverse collection of 23,500 segments exceeding 113 hours of data, representing over 1000 unique male and female speakers. Each audio file is provided in a single-channel 16kHz format, while the accompanying visual data boasts a frame rate of 30 fps. Furthermore, the development set (devset) contains 3,306 video recordings (approximately 8 hours), meticulously curated to ensure no speaker overlap exists between the train and dev sets, adhering to speaker independence criteria. To expand data diversity and augment the training process, we constructed a more challenging noisy dataset. This custom dataset meticulously integrates noise samples from three established sources: the Deep Noise Suppression Challenge (DNS) dataset \cite{reddy2021icassp}, the Diverse Environments Multichannel Acoustic Noise Database (DEMAND) \cite{thiemann2013diverse}, and the Clarity Enhancement Challenge (CEC1) \cite{graetzer2021clarity}. This enriched dataset further strengthens the model's ability to effectively enhance speech in a wider range of complex, real-world noise scenarios.
}
\textcolor{blue}{
To enhance the training process and expose the model to diverse noise scenarios, we curated a custom noise dataset from three distinct sources: CEC1, DEMAND, and DNS. This approach enriched our dataset with a variety of noise types, encompassing both domestic and multi-channel recordings. CEC1 provided 7 hours of everyday soundscapes, familiarizing the model with common environmental noises. DEMAND contributed 1 hour of recordings across 18 distinct soundscapes, further expanding the noise diversity through its multi-channel approach. Finally, DNS bolstered the dataset with 25 hours of unique noise selections sourced from Freesound, ensuring our model encountered a wide range of sounds beyond the other datasets.
}
\textcolor{blue}{
The combined noise dataset was carefully segmented into training (75\%), validation (15\%), and testing (10\%) sets. To simulate realistic noisy environments during training and validation, clean utterances were corrupted with random noise samples at varying Signal-to-Noise Ratio (SNR) levels, ranging from -9 to 6 dB. Importantly, noise types and SNR levels were kept distinct within each set to maintain the integrity and quality of our training data. For comprehensive testing, the clean test set was divided based on target SNR levels. Each clean utterance was then contaminated with a random noise sample from the dedicated test noise split, enabling systematic evaluation across diverse noise conditions. This approach ensured the model's ability to perform well in a wide range of realistic scenarios.
}
\textcolor{blue}{
To quantitatively assess the effectiveness of our proposed system in enhancing speech in noisy environments, we leverage several objective evaluation metrics: Perceptual Evaluation of Speech Quality (PESQ) \cite{rix2001perceptual}, Short-Time Objective Intelligibility (STOI)  \cite{taal2010short}, and Speech Distortion Index (SDI) \cite{hu2007evaluation}. These metrics provide valuable insights by offering numerical scores reflecting the system's ability to: Improve speech quality: PESQ (evaluates perceived speech quality, mimicking human judgment), Enhance intelligibility: STOI (measures the intelligibility of spoken words, quantifying how well the message is understood), Reduce noise: SDI (assesses the level of noise suppression achieved by the system).
}
\textcolor{blue}{
By analyzing these metrics under various noise conditions, we gain a comprehensive understanding of the system's performance and its potential for real-world applications.
}
\begin{comment}
To create noisy data for the train and devset in the first scenario, a speech utterance from the same corpus but from the disjoint set is chosen as speech interference and contaminated with the clean data, at random SNR levels between -9 and +6 dB. Next, a more difficult noisy dataset is created by combining noises from three different datasets, namely DNS, the DEMAND, and the CEC1 datasets. For DNS noises, we focused on sounds sourced exclusively from Freesound to ensure diversity. From the DEMAND dataset, we selectively included a single channel from each soundscape, carefully avoiding redundancies with Clarity Challenge sounds. Additionally, we excluded soundscapes categorized as "domestic" to prevent overlap with domestic environmental noises from CEC1. We also omitted the soundscape labeled as "OMEETING" due to its similarity to a competing speaker scenario. From the DNS dataset, we further refined our selections by filtering out sounds associated with the "Fan" category, thereby preventing duplications with Clarity Challenge sounds. 
\end{comment}

\section{Experimental Results}
\textcolor{blue}{
We first analyse how the integration of emotion as a contextual queue described in Section II influences the SE performance of the proposed DL-based AV SE system. In particular, we look at the sensitivity of the DL-based SE system for different types of background noises, over a wide range of SNRs to compare custom loss functions fairly in subsequent investigations. 
%The spectrogram and subjective listening tests are next analysed for further comparison.
%\subsection{Performance Improvement with Emotion Cues}
%In this section, we present and discuss the experimental results of our AV emotion-sensitive SE (Emotion-AVSE) system, compared to two baseline systems: the ASE and the AVSE system. 
For the initial comparison between the ASE and AVSE systems, we observe that the integration of visual cues into the ASE system significantly enhances speech quality, as evident in the higher PESQ scores. This improvement is attributed to the ability of visual information to guide the enhancement process and suppress noise more effectively. The AVSE system outperforms the ASE system, as evidenced by higher PESQ scores, indicating a noticeable increase in speech quality. Furthermore, the AVSE system exhibits better intelligibility, as evidenced by higher STOI scores, suggesting that the combined audio-visual model successfully leverages both modalities to enhance speech intelligibility. Consequently, the estimated speech signal experiences less distortion, as indicated by lower SDI scores.
}
\textcolor{blue}{
Building upon these findings, we incorporate emotional features as contextual cues for the SE process. This innovative approach further enhances the system's performance, leading to excellent results in terms of PESQ, STOI, and SDI. The integration of emotional cues significantly improves speech quality, evident in higher PESQ scores. Furthermore, the emotional-aware system achieves superior intelligibility, as indicated by elevated STOI scores, suggesting that emotional information aids in effectively enhancing speech intelligibility, even in extremely noisy conditions. Additionally, the system minimizes distortion in the estimated speech signal, as observed in reduced SDI scores. These results highlight the potential of emotional cues to further enhance the performance of audio-visual speech enhancement systems in challenging acoustic environments, ultimately contributing to more effective and context-aware communication solutions.
}
\textcolor{blue}{
Table 1 presents a performance comparison between ASE, AVSE, and our proposed Emotion-AVSE system (termed E-AVSE), against three baseline models: CochleaNet \cite{gogate2020cochleanet}, IO-AVSE \cite{hussain2021towards}, and LaiNet \cite{lai2023audio}. We evaluated these systems using standard l1-loss functions, except for IO-AVSE, which employed STOI as a loss function for parameter optimization. The performance analysis reveals that our AVSE system outperforms the baseline models, as indicated by better PESQ and STOI scores. This suggests that the integration of visual information into the audio-only framework significantly enhances speech quality and intelligibility. Furthermore, the AVSE system demonstrates substantial advantages, with marked improvements in terms of speech quality and intelligibility compared to the baseline systems. Although the integration of emotion information into the AVSE system slightly degrades speech quality, the resulting score remains competitive and, more importantly, improves speech intelligibility slightly. This implies that the Emotion-AVSE system effectively leverages emotion cues to maintain robust performance in terms of speech intelligibility, even in challenging noisy conditions. In summary, the results in Table 1 highlight the effectiveness of our system, which, despite a minor reduction in speech quality, excels in enhancing speech intelligibility, demonstrating its potential as a context-aware solution for SE in the presence of acoustic disturbances and emotional nuances.
}

\begin{table}[t]
%\normalsize
\setlength\tabcolsep{5pt}
%\captionsetup{justification=centering}
\caption{\scshape Performance comparison of proposed emotion-aware AVSE system with different Baseline models.}
\centering
\begin{tabular}{c|c|ccc}
\hline
\hline
\multirow{2}{*}{\textbf{Framework}}  & \multirow{1}{*}{\textbf{Loss}} & \multicolumn{3}{c}{\textbf{Objective Evaluation Metrics}} \\ \cline{3-5}

  & \multirow{1}{*}{\textbf{Function}} & \multicolumn{1}{c}{PESQ} & \multicolumn{1}{c}{STOI} & \multicolumn{1}{c}{SDI} \\ \cline{2-5} 
 
\textbf{Noisy} & -- & 1.326  & 0.447 & 3.061  \\ \hline
{UNet}\textsubscript{A} & MAE & 1.547 & 0.552 & 2.441    \\ \hline
CochleaNet & MAE & 1.633 & 0.588 & 2.014    \\
LaiNet \cite{lai2023audio} & MAE & 1.718 & 0.612 & 2.122    \\
IO-AVSE & STOI & 1.690 & 0.602 & 2.011    \\

\textbf{UNet}\textsubscript{E-AVSE (ours)} \ & MAE & 1.685 & 0.703 & 1.994    \\ 

\hline
\hline
\end{tabular}
%}
     \end{table}
%%%-----------------------------------------------------------------
\textcolor{blue}{
In Table 2, we present the results of optimizing our proposed model with three different loss functions: l1-loss, STOI loss, and modulation domain loss function \cite{vuong2021modulation}. These diverse loss functions offer multiple perspectives on the performance of our AVSE system integrated with emotion cues. Our findings reveal that, when optimized with STOI and modulation-domain loss functions, our system demonstrates superior performance compared to the l1-loss function. This outcome underscores the significance of considering various optimization strategies in the context of emotion-aware AVSE. The adoption of different loss functions serves as a pivotal step in our research, as it not only highlights the adaptability of our model but also underscores the flexibility of integrating emotion as a contextual cue. This multi-pronged approach enables us to fine-tune our model parameters and understand the intricacies of emotional cues for SE. Such findings are invaluable for the development of a robust emotion-aware AVSE system that can operate optimally across diverse acoustic conditions and enrich the landscape of affective computing.
}

\begin{table}[t]
%\normalsize
\setlength\tabcolsep{5pt}
%\captionsetup{justification=centering}
\caption{\scshape Performance comparison of proposed emotion-aware AVSE system using different loss functions.}
\centering
\begin{tabular}{c|c|ccc}
\hline
\hline
\multirow{2}{*}{\textbf{Framework}}  & \multirow{1}{*}{\textbf{Loss}} &  \multicolumn{3}{c}{\textbf{Objective Evaluation Metrics}}  \\ \cline{3-5}

  & \multirow{1}{*}{\textbf{Function}} & \multicolumn{1}{c}{PESQ} & \multicolumn{1}{c}{STOI} & \multicolumn{1}{c}{SDI}  \\ \cline{2-5}
 
\hline
\hline

\textbf{UNet}\textsubscript{E-AVSE} \ & MAE & 1.685 & 0.703 & 1.994    \\

\textbf{UNet}\textsubscript{E-AVSE} \ & STOI & 1.661 & 0.695 & 2.119    \\

\textbf{UNet}\textsubscript{E-AVSE} \ & Modulation-loss & 1.692 & 0.710 & 2.005    \\

\hline
\hline
\end{tabular}
%}
\end{table}
\section{Conclusion}
\textcolor{blue}{
By integrating emotion recognition into audio-visual speech enhancement (AVSE), our work presents a novel approach with demonstrably improved performance. Leveraging a U-Net architecture, our model significantly outperforms baselines, highlighting the value of incorporating emotional and visual cues. Integrating emotion recognition enables the system to adapt its processing based on the speaker's emotional state, resulting in context-aware and emotionally relevant speech enhancement. This adaptability is particularly impactful in challenging noise environments where traditional methods struggle.
}
\textcolor{blue}{
This emotion-based approach holds vast potential. It can advance assistive technologies for individuals with hearing impairments, enabling them to better perceive and understand emotional nuances in speech. Additionally, it can enhance conversational AI systems, facilitating more natural and engaging interactions. Furthermore, by preserving emotional cues crucial for human communication, our model paves the way for richer and more impactful interactions across various settings. This work signifies a transformative advancement, and the integration of emotions into speech enhancement promises significant societal benefits.
}

\bibliographystyle{IEEEtranS}
\bibliography{ref}

\end{document}